\documentstyle[epsfig,11pt]{article}

\def\edcomment#1{\iffalse\marginpar{\raggedright\sl#1\/}\else\relax\fi}
\reversemarginpar

\begin{document}
\title{Achieving {\em Extreme Resolution} in Numerical Cosmology Using 
Adaptive Mesh Refinement: Resolving Primordial Star
Formation\footnote{
Further information and
visualizations can be found at {\tt http://www.TomAbel.com/GB/}}}

\author{Greg L. Bryan\footnote{
Massachusetts Institute of Technology,
MIT 6-216, 77 Massachusetts Ave.,
Cambridge, MA 02139},
Tom Abel\footnote{
Harvard-Smithsonian CFA, Cambridge, MA 02138}, and
Michael L. Norman\footnote{
Lab. for Computational Asrophysics, UC San Diego, La Jolla, CA
92093-0424}
\footnote{Permission to make digital or hard copies of all or part of
  this work for personal or classroom use is granted without free
  provided that copies bear this notice and the full citation on the
  first page.  To copy otherwise, to republish, to post on servers or
  to redistribute to lists, requires prior specific permission and/or
  a fee.  SC2001 November 2001, Denver (c) 2001 ACM
  1-58113-293-X/01/0011 \$5.00}
}
\maketitle

\begin{abstract}
  
As an entry for the 2001 Gordon Bell Award in the ``special''
category, we describe our 3-d, hybrid, adaptive mesh refinement
(AMR) code {\em Enzo} designed for high-resolution, multiphysics,
cosmological structure formation simulations. Our parallel
implementation places no limit on the depth or complexity of the
adaptive grid hierarchy, allowing us to achieve unprecedented
spatial and temporal dynamic range. We report on a simulation of
primordial star formation which develops over 8000 subgrids at 34
levels of refinement to achieve a local refinement of a factor of
$10^{12}$ in space and time. This allows us to resolve the
properties of the first stars which form in the universe assuming
standard physics and a standard cosmological model.  Achieving {\em
extreme resolution} requires the use of 128-bit extended precision
arithmetic (EPA) to accurately specify the subgrid positions.  We
describe our EPA AMR implementation on the IBM SP2 Blue Horizon
system at the San Diego Supercomputer Center.

\end{abstract}


\section{Introduction}

Cosmic structure is formed by the gravitational amplification of
initially small density fluctuations present in the early universe.  A
fluctuation containing the mass of the Milky Way galaxy will collapse
by a factor of $\sim 10^3$ in size before it comes into dynamical
equilibrium. In order to adequately resolve its internal structure,
another two orders of magnitude of spatial resolution {\em per
dimension} are needed, at a minimum, for a total spatial dynamic range
$SDR=10^5$. Resolving the formation of individual stars in an entire
galaxy would require vastly more resolution: $SDR\sim 10^{20}$---a
seemingly unreachable goal.

N-body tree codes are widely used in numerical cosmology to achieve
high spatial dynamic ranges in general 3-d evolutions, and some
parallel implementations have won past Gordon Bell Awards
\cite{warren92,fukushige96,warren97,makino00}. The largest
cosmological N-body simulation used just over $10^9$ particles and
achieved a $SDR \sim 10^4$ \cite{virgo}. The highest dynamic range
N-body simulation achieved $SDR =2\times10^5$ with substantially fewer
particles \cite{springel00}. These calculations simulate only the
collisionless cold dark matter (CDM) which dominates the gravitational
dynamics of structure formation.

We are interested in simulating the formation of cosmic structures
including the all--important baryonic gas which forms the visible
galaxies. Our original design goal was to achieve $SDR=10^4$, a mark
which we have far surpassed. In this paper we describe the Enzo
cosmological adaptive mesh refinement (AMR) code we have developed for
parallel computers and its application to a simulation of the
formation of the first stars in the universe.  The time-dependent
calculation is carried out in full 3-d on a structured adaptive grid
hierarchy which follows the collapsing protogalaxy and subsequent
protostellar cloud to near stellar density starting from primordial
fluctuations a few million years after the big bang. More than 8000
subgrids at 34 levels of refinement are generated automatically to
achieve a final spatial and temporal dynamic range of $10^{12}$ (for
comparison, $10^{12}$ is roughly the ratio of the diameter of the
earth to the size of a human cell).
%Temporally, $10^{12}$ is roughly
%the ratio of time since the extinction of the dinosaurs to when you woke
%up this morning. 
It is the highest dynamic range 3-d simulation ever
carried out in astrophysics.

The calculation incorporates all known relevant dynamical, chemical,
and thermodynamic processes and is carried out in a proper expanding
cosmological background spacetime. Enzo combines an Euler solver for
the primordial gas \cite{bryan95}, an N-body solver for the
collisionless dark matter \cite{bryan98}, a Poisson solver for the
gravitational field \cite{norman98}, and a 12-species stiff reaction
flow solver for the primordial gas chemistry \cite{aazn97}. The latter
is needed to determine the nonequilibrium abundance of molecular
hydrogen which dominates the radiative cooling of the gas. This system
of equations is solved on every level of the grid hierarchy with full
two-way coupling, making it one of the most complex AMR simulations
ever carried out.  Adaptive mesh refinement allows us to resolve
locally all the important length and timescales everywhere within the
gravitationally collapsing cloud at all times, giving us confidence
that our results are accurate. The calculation is effectively an {\em
ab initio} simulation of star formation which connects initial
conditions to the final state in full generality.

In this paper, we concentrate on the technical and performance
aspects of the AMR code, but also discuss some of the exciting
astrophysical results we have obtained.  As we will see, achieving
extreme resolution is more a matter of high performance data
structures and extended precision arithmetic than raw gigaflops,
although the latter is certainly needed.

% ==============================================================

\section{Physical Model}

The physics of present-day star formation is complicated and fraught
with uncertainties.  This is due to the relative complexity of the
interstellar medium: it is composed primarily of hydrogen and helium
but also contains heavier elements which are important for radiative
cooling and heating.  In addition, there are dynamically important
magnetic fields, radiative backgrounds and uncertain initial
conditions.  

In contrast, the formation of the first star takes place in a much
simpler environment: the gas composition is essentially just hydrogen
and helium (since heavier elements are formed inside stars), there is
no important radiative source, the magnetic fields are most likely
insignificant and --- most importantly --- the initial conditions are
precisely specified by cosmological models.

This makes primordial star formation a clean initial value problem in
which we can include, from first principles, all of the relevant
physical processes.  Moreover, the problem is extremely important --
it provides the starting point for the formation for all other
structure in the universe, from galaxies to superclusters.  The
combination of all of these elements is rare in physics, and efforts
to solve the problem date back over three decades.  However, progress
has been difficult because of the extreme multi-scale nature of the
problem, as discussed in the introduction.  In the following, we
review the physical ingredients which are required to solve this
problem, emphasizing the numerical implications.

\subsection{Initial Conditions}

In modern cosmological models, cold dark matter (CDM) dominates the
dynamics of structure formation because its mass density exceeds that
of the ordinary baryonic material by a factor of ten or more.
Although the composition of CDM is unknown, its cosmologically
relevant properties---its mean density and the functional form of its
power spectrum of density fluctuations $P(k)$---are known or are
calculable once a Friedmann ``world'' model is specified.  We simulate
a ``standard'' CDM model which means that the cosmological parameters
and the amplitude of the power spectrum are those that have been shown
to reproduce the statistical properties of galaxies and clusters of
galaxies in the present universe \cite{ostriker93}.

The CDM power spectrum has the interesting property that {\it rms}
density fluctuations are logarithmically divergent on small mass
scales, implying that structures form hierarchically from the
bottom-up: small gravitational condensations form first which then
merge to form larger, more massive, objects later on.  This merging
behavior is captured numerically by combining an N-body technique, to
follow the motion of the collisionless dark matter particles, with a
Poisson solver, to compute the self-gravity of the dark matter (and
the gas).  But when does collapse to high density occur?  How does it
proceed? And what kind of astrophysical object is produced?

\subsection{Gas Physics and Chemistry}

In order to answer these questions, we need an accurate description of
the thermal state of the gas.  The fluid equations plus the ideal gas
law (and Poisson's equation) describe the evolution of the gas
density, velocity and energy, but require knowledge of the radiative
heating and cooling rates.  This, in turn, depends on the chemical
composition.

The primordial gas contains about 76\% hydrogen and 24\% of helium
(trace amounts of deuterium and lithium play no important role here).
For such a gas, and for temperatures below $10^4$ K, the primary
cooling agent is molecular hydrogen, which has rotational and
vibrational excited states than can be collisionally excited.  Because
the cosmological background density of baryons is small, chemical
reactions in the smooth background gas occur on long timescales. As a
consequence chemical equilibrium is rarely an appropriate assumption.
The number of possible chemical reactions involving H$_2$ is large,
even in the simple case of metal free primordial gas.  We have
tediously selected the dominant reactions and collected the most
accurate reaction rates available \cite{pgas97}.  We solve the time
dependent chemical reaction network involving twelve species
(including deuterium and helium).  A fast numerical method to solve
this set of stiff ordinary differential equations has been developed
by some of us \cite{aazn97}.

Once the chemical state is known, we must determine the radiative
cooling rate.  For densities encountered in the collapsing cloud, it
is accurate to assume the gas is optically thin and only consider
excitations from atomic and molecular ground states.  We include all
known radiative loss terms due to atoms, ions, and molecules that are
appropriate for our primordial gas.  Also the energy exchange between
the cosmic microwave background and free electrons (Compton heating
and cooling) is included.

\section{The Enzo Cosmological Adaptive Mesh Refinement Code}

To solve the star formation problem using a traditional static,
uniform mesh is simply not feasible.  Even assuming a continuation of
Moore's law in its present form, it would not be until about 2200 that
a problem of this dynamic range could even fit into memory of the
largest systems.  The solution method that we adopt is structured
adaptive mesh refinement (SAMR: \cite{berger89}), because it: (1) is
spatially- and time-adaptive, (2) uses accurate and well-tested
grid-based methods for solving the hydrodynamics equations, and (3)
can be well optimized and parallelized.  In this section, we describe
our implementation; more details are available elsewhere
\cite{bryan98,norman98,bryan99}.

\subsection{Structured Adaptive Mesh Refinement}

The central idea behind SAMR is deceptively simple.  While solving the
desired set of equations on a uniform grid, monitor the quality of the
solution; when necessary, add an additional, finer mesh over the
region that requires enhanced resolution.  This finer (child) mesh
obtains its boundary conditions from the coarser (parent) grid or from
other neighbouring (sibling) grids with the same mesh spacing.  The
finer grid is also used to improve the solution on its parent.  As the
evolution continues, it may be necessary to move, resize or remove the
finer mesh.  Even finer meshes may be required, producing a tree
structure that may continue to any depth.

\begin{figure}
\epsfxsize=5in
\centerline{\epsfbox{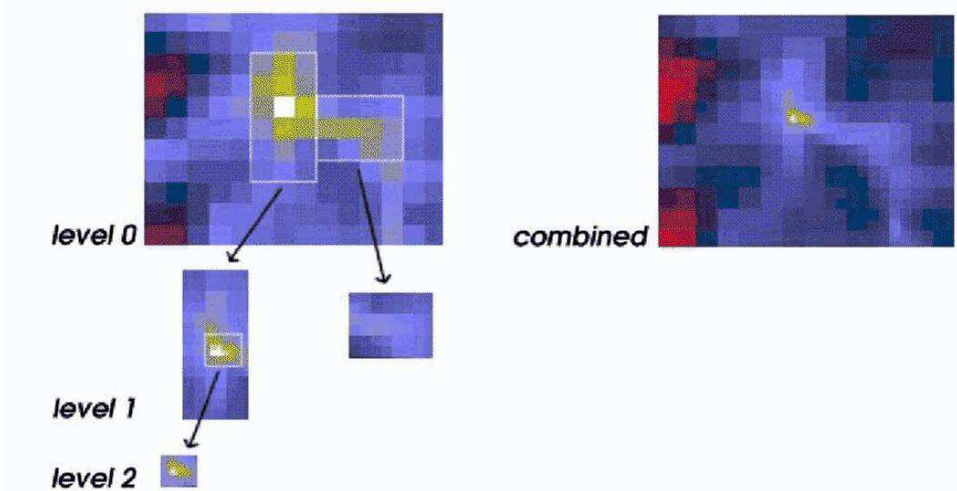}}
\caption{In this two-dimensional example of SAMR, a root grid has two
sub-grids with one-half the mesh spacing and one sub-grid has an
additional sub-sub-grid with even higher resolution.  The tree
structure on the left represents how these data are stored, while on
the right we show the resulting composite solution.}
\label{fig:amr_overview}
\end{figure}

We show an example of this for the two-dimensional case in
Figure~\ref{fig:amr_overview}.  Each level up the hierarchy decreases
the mesh spacing relative to its parent by a factor $r$ (the
refinement factor).  Therefore, if we define the resolution (or SDR)
as the ratio of the length of the entire region to the cell spacing,
the resolution at a level $l$ is given by $nr^l$, where $n$ is the
number of cells in the root grid.

There are a number of things to note about this algorithm.  First, in
order to be efficient, the covering fraction of the refined meshes
must be relatively small.  For our problem the covering fraction of
the most refined grids is extremely small, approximately $10^{-30}$
(nevertheless, this is by far the most interesting region!).  Second,
the refinement factor is constrained to be an integer so that meshes
can be aligned, making communication between grids easier and more
accurate.  In addition, the subgrids must be rectangular and
completely contained within their parents.  Third, the finer mesh is
in addition to, not instead of, the coarse mesh it covers.  At first
this seems wasteful; however, the overhead is quite small and there
are significant advantages.  The individual grids remain uniform, so
we can use off-the-shelf solvers.  Also, since the timestep constraint
depends on the mesh spacing, the uniform (coarse) mesh has a uniform
(large) timestep, thus naturally permitting multiple timesteps at
different levels.

\subsection{Advancing the Grid Hierarchy}

\begin{figure}
\epsfysize=3in
\centerline{\epsfbox{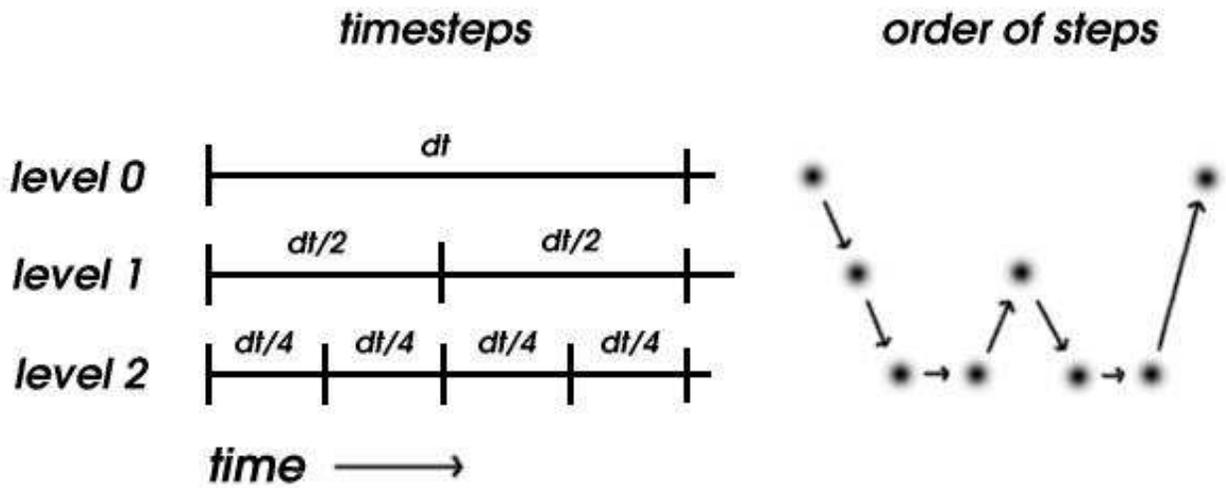}}
\caption{This shows the order of timesteps for the SAMR example
  depicted in Figure~\ref{fig:amr_overview}.  
  First the root grid is advanced,
  and then the subgrids ``catch-up''.  This permits the calculation of
  time-centered subgrid boundary conditions for higher temporal
  accuracy.}
\label{fig:timesteps}
\end{figure}

To advance our system of coupled equations in time as represented on
this hierarchy requires some care.  The overall control algorithm is
quite similar to that used for a single grid.  The heart of the
algorithm is the recursive {\tt EvolveLevel} routine which is passed
the level it is to work on and the new time:
\begin{verbatim}
EvolveLevel(level, ParentTime)
begin
  SetBoundaryValues(all grids)
  while (Time < ParentTime)
  begin
    dt = ComputeTimeStep(all grids)
    SolveHydroEquations(allgrids, dt)
    Time += dt
    SetBoundaryValues(all grids)
    EvolveLevel(level+1, Time)
    FluxCorrection
    Projection
    RebuildHierarchy(level+1)
  end
end
\end{verbatim}
The interior loop is used to advance the grids on this level from {\tt
Time} (the time at their current state) to {\tt ParentTime} (the
time to which the parent grids have been advanced).  Inside this loop
there is a recursive call so that all finer levels
are advanced as well.  The resulting order of timesteps is
like the multigrid W cycle, as shown in Figure~\ref{fig:timesteps}.

\subsubsection{Solving the equations of hydrodynamics}

We treat each grid as its own self-contained unit, so that each
routine in the pseudo-code above is applied to all the grids on a
given level.  As with any hyperbolic equation, we first need to set
the boundary conditions on the grids.  For the root grid, we use a
predefined boundary condition (periodic in this case), while for
finer grids, a two step procedure is used:
\begin{enumerate}
\item All boundary values are first interpolated from the grid's
  parent.
\item Grids which border other grids on the same level (i.e. siblings)
  use the solution from the sibling grid.
\end{enumerate}
This ensures that all boundary values are set using the highest
resolution solution available.

Once the boundary values have been set, it's straightforward to solve
the equations of hydrodynamics on a grid.  Because it is a Cartesian
rectangular grid, we can use highly-optimized solvers.  We have
implemented two: the piecewise parabolic method (PPM) \cite{ppm}
modified for use in cosmological hydrodynamics \cite{bryan95}, as well
as a robust finite difference technique \cite{zeus}.  This allows us
a double check on any result.

A disadvantage to such a hierarchy is that we must ensure consistency
in regions which are represented by both coarse and fine meshes.  This
involves two steps: the first is to correct the coarse fluxes (of
conserved quantities) at subgrid boundaries to reflect the improved
flux estimates from the subgrid.  This is required to ensure mass,
momentum and energy conservation as material flows into and out of a
refined region.  The second step, termed {\it projection}, updates the
solution on the coarse mesh points which are covered by finer meshes.
Taken together, these two steps represent one side of the two-way
communication between parent and child grids (the other side being
boundary interpolation from parent to grid).

\subsubsection{Rebuilding the Grid Hierarchy}

The final part of the {\tt EvolveLevel} routine adapts the grid
hierarchy to the new solution.  The {\tt RebuildHierarchy} procedure
modifies (``rebuilds'') the grids on the specified level and higher.
This involves three steps:
\begin{enumerate}
\item A refinement test is applied to the parent grids of the current
  level, resulting in a boolean field indicating if this mesh point
  needs to be refined.
\item Rectangular regions are chosen which cover all of the refined
  regions, while attempting to minimize the number of unnecessarily
  refined points.  This is done with an edge-detection algorithm
  from machine vision studies \cite{berger91}.
\item Finally, the new grids are created and their values are copied
  from the old grids on the same level where possible, or are
  otherwise interpolated from the parent grids.  The old grids are
  then deleted, freeing their memory.
\end{enumerate}
This process is repeated on the next refined level, until the entire
hierarchy has been rebuilt.  This is both easier and more
effective than moving and resizing the grids, and still only takes
about 10\% of the overall cpu time.

\subsubsection{Refinement Criteria}

In this application, there are three criteria used to determine when
refinement is required.  Since gravitational collapse causes mass to
flow into a small number of cells, the first two are designed to
preserve a given mass resolution in the solution.  We define a
threshold $M_{*}$ such that whenever a cell accumulates at least this
much mass, it is refined.  This is done separately for the gas and the
dark matter component.

The third refinement criterion ensures that all possible collapsing
fragments are resolved.  More specifically, we require that the cell
width be less than some fraction of the local Jeans length ($\Delta x
< L_J/N_J$), the smallest perturbation which is unstable to
gravitational collapse.  We have varied $N_J$, the number of cells
across the local Jeans length, from 4 to 64 without seeing a
significant difference in the results.

\subsection{Adding Physics: Gravity, Dark Matter and Chemistry}

To simulate a system such as a collapsing star, we must also solve the
elliptic Poisson equation on the hierarchy.  On the root grid, this is
done with an FFT which naturally provides the periodic boundary
conditions required.  On subgrids, we interpolate the gravitational
potential field and then solve the Poisson equation using a
traditional multi-grid relaxation technique.  In order to produce a
solution that is consistent across the boundaries of sibling grids, we
use an iterative method: first solving each grid separately,
exchanging boundary conditions, and then solving again.

In contrast to the familiar baryonic fluid, the dark matter is
pressureless and collisionless, only interacting via gravity.
Therefore, it is not appropriate to use the fluid equations; rather,
we solve for the individual trajectories of a representative sample of
particles.  This we do using particle-mesh techniques specially
tailored to adaptive mesh hierarchies.  This adds another layer of
complexity to the software, the details of which are discussed
elsewhere \cite{bryan98}.

Finally, as discussed earlier, the cooling of the primordial star is
controlled by the formation of molecular hydrogen.  Therefore, we need
to follow the chemical evolution of various ionized states of
hydrogen, helium, deuterium, and molecular hydrogen, twelve in all.
This coupled set of ordinary differential equations can be solved
independently for each grid point.  Because the equations are stiff,
we use a backward finite-difference technique for stability,
sub-cycling within a fluid timestep for additional accuracy
\cite{aazn97}.

\subsection{Parallelization Strategy}

The code is implemented in C++, with compute-intensive kernels in
FORTRAN 77.  An object oriented approach provides a number of
benefits.  The first is encapsulation: a grid represents the basic
building block of AMR.  We have defined a series of atomic and binary
operations that perform functions such as solving a certain set of
equations on a grid or interpolating from one grid to another.  This
greatly simplifies the control algorithms.  The second is
extensibility: adding new equations is generally a matter of adding a
few methods to a grid class and a few lines to the control algorithm.

Efficiently parallelizing SAMR is difficult, particularly for
distributed memory systems.  Grids have a relatively short
life, so information must be updated frequently.  Moreover, load
balancing becomes a serious headache since small regions of the
original grid eventually dominate the computational requirements.

As a first attempt, we tried compiler-based parallelization (a
predecessor to OpenMP).  Using pragmas, we identified large-grained
loops that performed operations over entire grids.  This was
successful for a relatively small number of processors (8-16), but
could not be extended to distributed memory systems, such as clusters
of PCs.  To adapt to these machines, we developed an MPI version that
exploited the following optimization techniques:

\begin{itemize}

\item {\em Distributed objects}.  We leveraged the object-oriented design by
distributing the objects over the processors, rather than attempting
to distribute an individual grid.  This makes sense because the grids
are generally small ($\sim 20^3$) and numerous (sometimes in excess of
50,000 grids).  This had the added benefit of ease of coding: we
simply added a small number of communication primitives to the class
methods.
  
\item {\em Sterile objects}.  Although distributing the objects results in
good load balancing, it can greatly increase the amount of
communication since each processor has to probe other processors to
find potential neighbours.  We solved this problem by creating a type
of object which contained information about the location and size of a
grid, but did not contain the actual solution.  These sterile objects
are small and so each processor can hold the entire hierarchy.  Only
those grids which are local to that processor are non-sterile.  This
means that almost all messages are direct data sends; very few probes
are required.
  
\item {\em Pipeline communications}.  One result of distribution is that all
binary operations (e.g. obtaining boundary values) are potentially
non-local.  However, a given stage of the calculation generally
involves many sends.  We optimize this by dividing each stage into two
steps.  First, all of the data (such as boundary values) are processed
and sent.  Since all processors have the location of all other grids
locally (thanks to the sterile objects), we can order these sends such
that the data that are required first are sent first.  Then, in the
receive stage, the data needed immediately have had a chance to
propagate across the network while the rest of the sends were
initiated.  This was implemented through the use of MPI's asynchronous
communication mode, and resulted in a large decrease in wait times.

\end{itemize}

\subsection{Extended Precision Arithmetic}

Performing a simulation with such extreme dynamic range presents
unique difficulties.  One of these is simply the requirement of being
able to accurately describe the positions of grids and particles
within the problem domain.  At a minimum, we must be able to
distinguish between the two locations $x + \Delta x$ and $x$, where
$\Delta x$ is the minimum grid spacing.  This implies a precision of
at least $\Delta x/x \sim 10^{-12}$ (in fact we require roughly 100
times larger than this because of various mathematical operations
which are applied to this ratio).  Therefore, we need 128 bit
precision.  Unfortunately, this is not always available and when
available, we often suffer a large performance penalty.  For example,
the GNU FORTRAN compilers does not support 128 bit precision.  The SGI
compiler on the Origin2000 does provide 128 bit arithmetic, but is
some 30 times slower than 64 bit.  The RS/6000 compiler on the SP2
allows 128 bit precision, but only as an extra compiler option, so
libraries (such as MPI) must be re-compiled.

One option for dealing with this issue is to develop routines which
use native 64 bit arithmetic to perform 128 bit operations,
circumventing patchy native support.  Recent work \cite{bailey} in
this direction appears to be quite successful and we are examining the
possibility of adopting this method.

Whichever approach is taken, it is clear that the use of 128 bit
arithmetic should be kept to a minimum because of its impact on
operation count, memory consumption and interprocessor message size.
Throughout the code, we have identified only those operations which
require high precision.  These correspond to operations involving
absolute position and time, as opposed to relative location (O($\Delta
x$)).  Most grid atomic operations can be entirely written in terms of
relative position and time, and hence low precision.  This reduced the
total high-precision operation count to $\sim 5\%$ of the total,
resulting in considerable speed (and memory) improvements.

% ==============================================================

\section{Astrophysical results}

In the calculation discussed in this paper, we have set up a three
dimensional volume that is 256 comoving kiloparsec (kpc) on a side (1
parsec is 3.26 light years, roughly the distance to the Sun's nearest
neighbour).  We set up the initial conditions a few million years
after the big bang, using an inflation-inspired cosmological model.
We first run a low-resolution ($64^3$) simulation to determine where
the first star will form and then restart the calculation including
three additional levels of static meshes, covering the region from
which the star will form (equivalent to $512^3$ initial conditions
over the entire box).  This allows us to increase mass resolution in
this region by a large factor (512) and capture as many
small-wavelength modes in the initial conditions as possible.  We have
experimented with using only two additional levels and find it has
little effect on the overall evolution of the object.  We have also
carried out a number of experiments varying the refinement criteria
and find the results described here are quite robust.

A series of earlier AMR simulations at increasing resolutions have
revealed the following results.  A simulation with
$SDR=2.5\times10^{5}$ \cite{abn00} showed that at redshifts of $\sim
20$ (approximately 150 million years after the big bang) primordial
gas becomes gravitationally bound to dark matter condensations with
characteristic mass of few $10^5$ M$_{\odot}$ (solar masses).
Non-equilibrium chemical reactions form $H_2$ from primordial hydrogen
in the cores of these clumps in sufficient concentrations to cool the
gas down to a few hundred K via molecular line radiation. A simulation
with $SDR=3\times10^{7}$ \cite{norman00} showed the onset of
gravitational instability in this cool gas and the formation of a
collapsing protostellar core. The mass of the collapsing cloud was a
few hundred solar masses. Most recently, a simulation with
$SDR=10^{10}$ \cite{abn01} resolved the supersonic accretion flow onto
a central, fully molecular protostellar core, confirming the resulting
star would be much more massive than the Sun.

\begin{figure}
\epsfxsize=6in
\centerline{\epsfbox{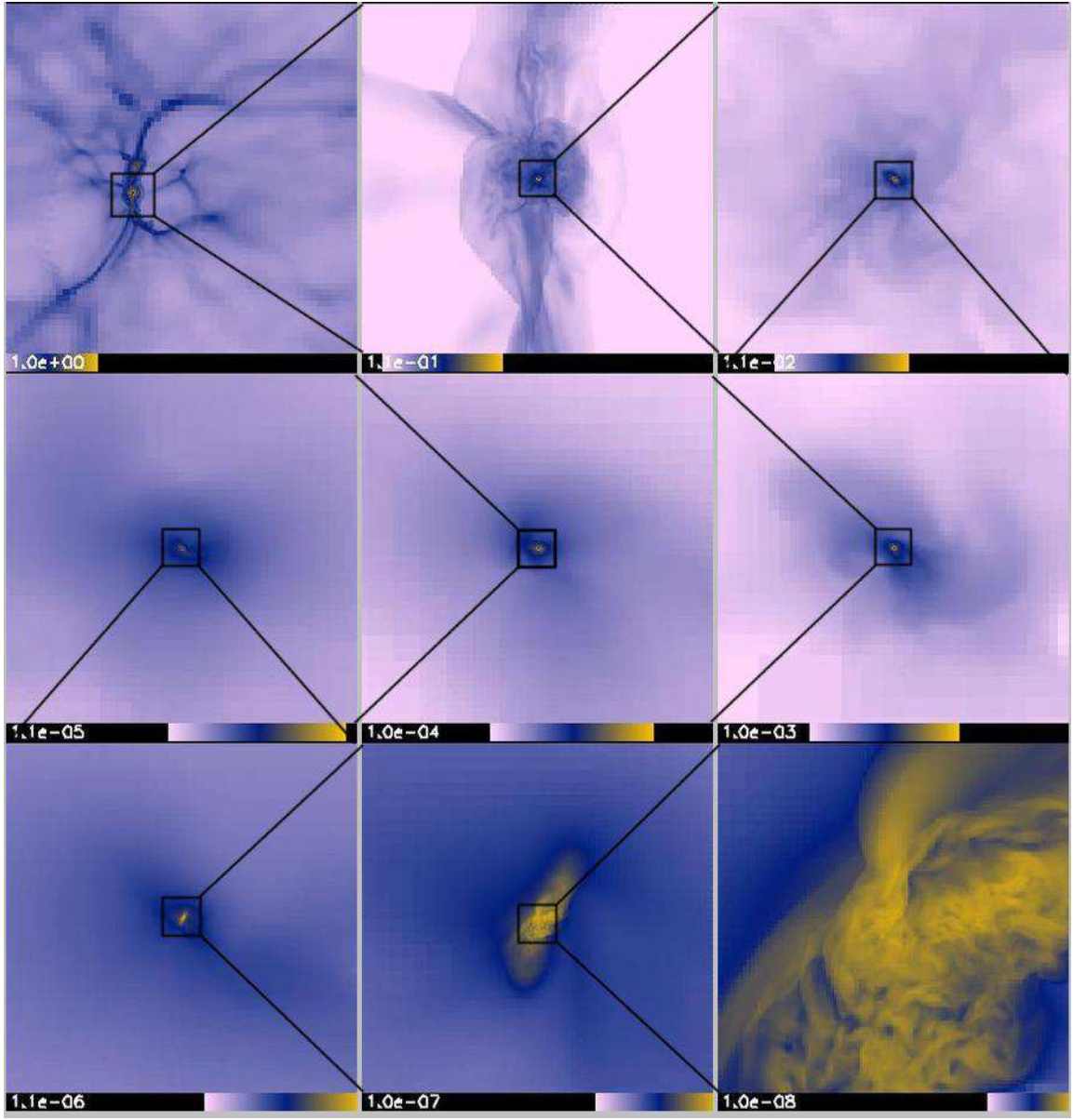}}
\caption{In these frames we show a zoom into the star forming region.
Each panel shows a slice of the logarithm of the gas density
magnified by a factor of ten relative to the previous frame,
starting at the upper left.}
\label{fig:density_zoom}
\end{figure}

The simulation reported here attempts to follow the protostar's
collapse to stellar density, which we estimate will require
$SDR=10^{15}$. As of current writing, we have achieved $SDR=10^{12}$
We find a rapidly accreting protostar that will form a single massive
star with a mass of order one hundred times that of our own sun.  Such
a large star will explode as a supernovae, generating the very first
heavy elements in the universe.  In Figure~\ref{fig:density_zoom}, we
show a zoom into the forming star.  At this stage, the protostar is
still collapsing and not yet spherical since gravity still overwhelms
internal pressure.  We highly encourage readers to
view the animated version of this zoom at {\tt
www.TomAbel.com/GB/movies.html}.

\begin{figure}
\epsfysize=11cm
\vspace{-.3cm}
\centerline{\epsfbox{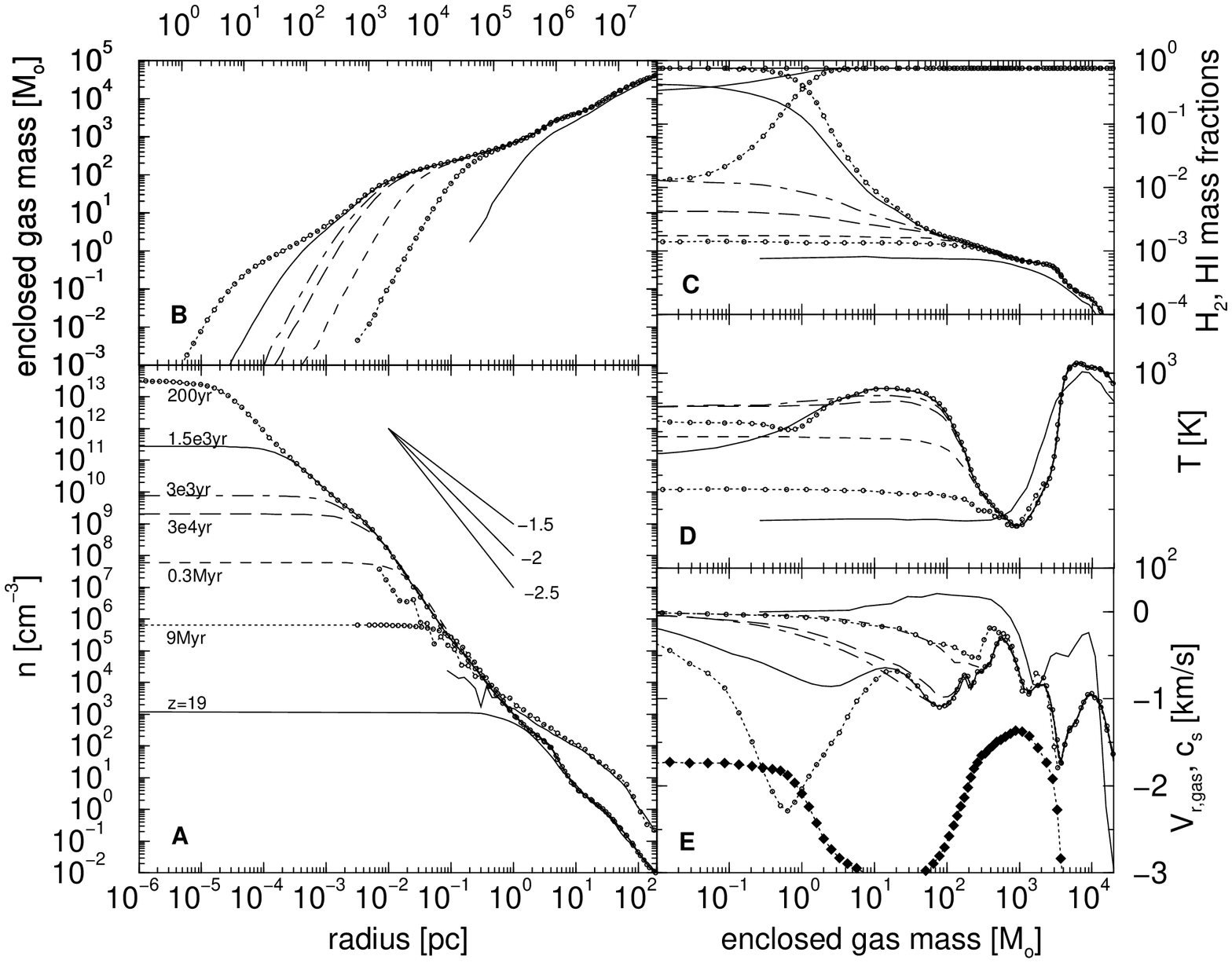}}
\caption{ Radial profiles of mass-weighted spherical about the densest
  point in the cloud of various
    physical quantities at seven different output times. Panel A shows
    the evolution of the particle number density in cm$^{-3}$ as a
    function of radius at redshift 19 (solid line), nine Myr later
    (dotted lines with circles), 0.3 Myr later (dashed line), $3
    \times 10^4$ years later (long dashed line), $3 \times 10^3$ years
    later (dot--dashed line), $1.5 \times 10^3$ years later (solid line) and
    finally 200 years later (dotted line with circles).
    The two lines between $10^{-2}$ and 200 pc give the
    DM mass density in GeV cm$^{-3}$ at z=19 and the final time,
    respectively. Panel B gives the enclosed gas mass as a function of
    radius. In C the mass fractions of atomic hydrogen and molecular
    hydrogen are shown. Panel D and E illustrate the temperature
    evolution and the mass weighted radial velocity of the baryons,
    respectively. The bottom line with filled symbols in panel E shows
    the negative value of the local speed of sound at the final time.
    In all panels equal output times correspond to equal line styles.
    The upper x--axis in panel B gives the radius in astronomical
    units. }\label{fig:five}
\end{figure}

Although the cloud and protostar are not spherical, it is instructive
to plot in Figure~\ref{fig:five} radial profiles of mass-weighted
spherical averages of various quantities for seven illustrative times
of the evolution.  At the first output time shown (marked z=19), about
700 $M_{\odot}$ have cooled below 200 Kelvin at the center of the
protogalactic halo of mass $5.44 \times 10^5 M_{\odot}$. From the
radial velocity profiles it is evident that cooling gas behind the
accretion shock (at $r \sim 100$ pc) falls onto this cooled central
material.  This cold central region within $r \sim 10$ pc is the high
redshift analog of a present day molecular cloud despite its minute
molecular mass fraction of $\sim 10^{-3}$ (see panel C of
Figure~\ref{fig:five}).

As the evolution continues (the second output time shown in
Figure~\ref{fig:five}) the total mass of the proto-galactic object has
grown to $7.4 \times 10^5 M_{\odot}$ at $z=18.209$.  The primordial
``molecular cloud'' contains now $\sim 1000 M_{\odot}$.  At the same
time a central core of $\sim 100M_{\odot}$ starts to become warmer
($\sim 300$ K) within the ``molecular cloud'' and shows small radial
velocities. For this central core the radial velocities and
temperatures are seen to increase monotonically as time goes on.  The
molecular mass fraction also starts to increase slightly. When the
central densities exceed $10^9$ cm$^{-3}$ three--body H$_2$ formation
shows its effect in binding atomic hydrogen to molecules. At central
densities $\sim 10^{11}$ cm$^{-3}$ atomic and molecular hydrogen exist
in similar abundance. This enhanced molecular fraction which decreases
the cooling time is the reason for the observed central temperature
drop (see panel D of Figure~\ref{fig:five}).  Interestingly, the inner
$\sim 1 M_{\odot}$ shows similar thermal evolution and start to
collapse more rapidly than its surrounding. In fact as can be seen in
panel E of Figure~\ref{fig:five} it eventually starts to contract
faster than the local sound speed. Before this dynamical collapse all
the cooling gas has been settling to the center with sound crossing
times shorter than or equal to the dynamical times, and long cooling
times. I.e. the contraction proceeds quasi--hydrostatically until 3
body H$_2$ formation turns the gas fully molecular. During this slow
contraction phase sound waves damp sub--Jeans density perturbations in
the baryons yielding a rather smooth gas distribution prior to the
onset of 3 body H$_2$ formation.  The time-scales become as short as
weeks for the densest material followed in the simulations. We have
not observed any signs of fragmentation by examining the
three--dimensional density field of the final output time.
Although the protostar has not yet reached stellar density, we
conclude that a single massive star will ultimately form.

\section{Code Performance Results}

Our primary interest is not raw performance, but rather the complexity
and difficulty of the problem and the algorithm.  Therefore, we first
examine some characteristics of the resulting grid hierarchy.

\begin{figure}
\epsfxsize=5in
\centerline{\epsfbox{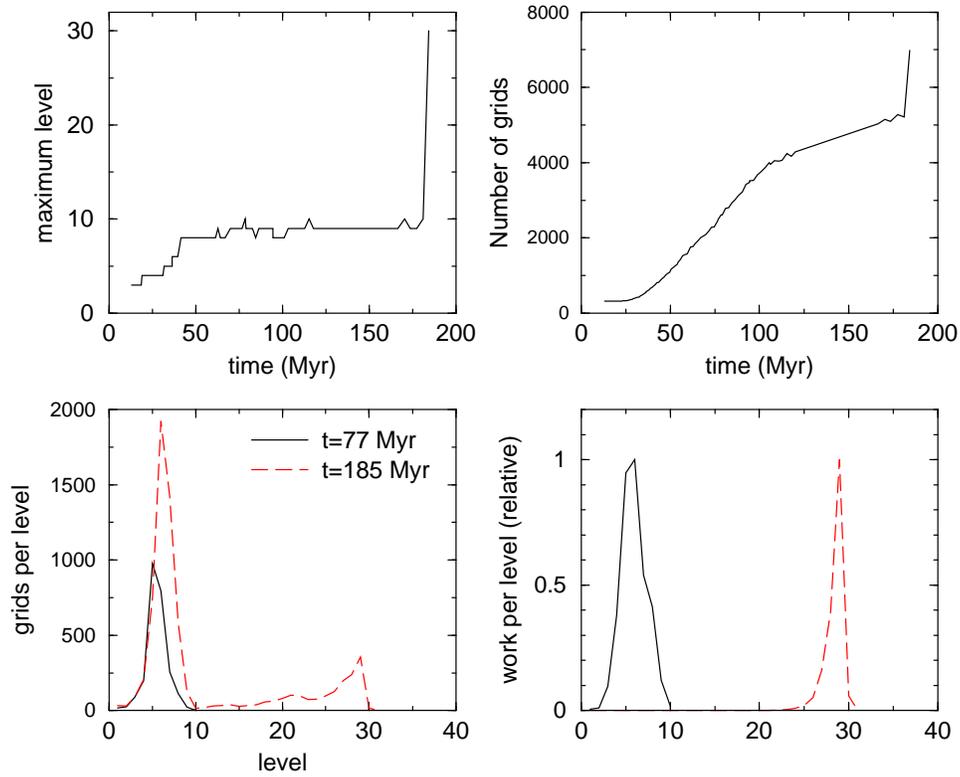}}
\caption{The top left and right panels show the depth of the hierarchy
  tree and the number of grids as a function of time (in millions of
  years),  The bottom left and right panels plot the number of grids 
  per level and an estimate of the computational work required per
  level (in each case normalized so that the maximum value is unity).}
\label{fig:grids}
\end{figure}

In the top two panels of Figure~\ref{fig:grids}, we show how the grid
hierarchy grows as time progresses.  Note the slow increase in the
number of grids as the proto-star condenses and the final, very sudden
jump in the depth of the grid tree at the end when the core of the
cloud collapses to high density.  This demonstrates how the data
structures themselves adapt to fit the physical solution.  Note also
the extremely large number of memory allocations and frees, since the
entire grid hierarchy is rebuilt thousands of times.  This kind of
method represents a new class of scientific computing that place great
strain on the operating system infrastructure.  Total memory usage is
also substantial, often reaching up to 20 GB.  With outputs in the 2-4
GB range, we require at least 50-100 GB disk storage and much more
mass storage space.

In the bottom two panels of this figure, we have chosen two
representative times and plotted the distribution of grids per level.
At early times, most of the grids are at moderate levels, representing
the fact that relatively low resolution is sufficient to model the
proto-stellar cloud.  However, at late times, as the proto-star forms,
a large investment is required at the very highest levels of
resolution.

Finally, we turn to the raw performance of the code.  Because of the
adaptive nature of the algorithm it is difficult to determine an
over-all flop rate for the entire computation.  One solution to this
problem would be to instrument each module to return its operation
count (which depends on such factors as the grid size, particle
number, and other parameters, as well as the solution itself in the
case of some modules).  Then we could simply add up the count returned
by all the modules.  However, the code is nearly 100,000 lines, so
this remains a future project.

In the meantime, we have estimated the flop rate in the following way.
We have used the hardware floating-point counter on the SGI Origin2000
to determine the total floating point operation count for a portion of
the simulation (drawn from near the middle of the run, which should be
representative).  In doing so, we have been careful to use only 64 bit
precision, rather than 128 bit (this somewhat affects the accuracy of
the result but because it is only used for timing this is not a
concern).  Then, when we run the calculation on our production
machine, the Blue Horizon SP2 system at the San Diego Supercomputer
Center, we use full 128 bit precision and measure the wall clock time
for this same portion of the simulation.  This ensures that the
operation count employs a widely-used standard (the hardware counter
on the R10000 processor), while the timing reflects the use of 128 bit
precision.  This provides a conservative estimate of the flop rate
since we treat 128 bit operations as a single operation, even though
the hardware itself may not do so.  Given this operation count
coupled with the measured wall clock time running on 64 processors of
the Blue Horizon SP2, produced a total speed of approximately 13
Gflop/s.  Since most of the run was done on 64 processors and the
section timed is representative of the entire calculation, this
represents our estimate of the sustained performance.

As an exercise, we can also ask how long this calculation would have
taken with a traditional static grid code and compute an effective or
virtual flop rate.  To do this, we assume a grid with $10^{12}$ cells
on each side, and assume the entire calculation would have taken
(quite conservatively) $10^{10}$ timesteps.  This works out to
approximately $10^{50}$ floating point operations.  Since the entire
calculation took of order $10^6$ seconds, this converts to a virtual
flop rate of $10^{44}$ flop/s.

We stress that this result has been achieved by optimizing all stages
of the calculation.  Since we solve a number of different kinds of
equations with various techniques, it is instructive to examine the
relative time taken by each part.  In the table below, we examine the
fractional part of the compute time taken by each of the following
science components: (a) the hydrodynamics solver, (b) the gravity
solver, (c) the chemistry solver, (d) the particle solver for the dark
matter, as well as a few major parts of the amr structure.  We stress
that these are fractions of the compute time which, for our 64
processor run, represents approximately 60\% of the total wall clock
time, the other 40\% being consumed by communication and imperfect
load balancing.  More detailed analysis of this code (on other
problems) has been performed elsewhere \cite{zlan01a}.

\begin{center}
\vspace{0.5cm}
\begin{tabular}{|l|c|}
\hline
component & usage \\  \hline \hline
hydrodynamics & 36 \% \\ \hline
Poisson solver & 17 \% \\ \hline
chemistry \& cooling & 11 \% \\ \hline
N-body & 1 \% \\ \hline \hline
hierarchy rebuild & 9 \%  \\ \hline
boundary conditions & 15 \% \\ \hline
other overhead & 11 \% \\ \hline
\end{tabular}
\vspace{0.5cm}
\end{center}

\section{Analysis and Visualization Challenges}

Beyond simply running the simulation, we have also faced major
challenges in data analysis and visualization of these large dynamic
range simulations. Our analysis routines have had to be implemented in
parallel to hold the gigabytes of data in memory on available systems.
These routines facilitate finding collapsed objects and other regions
of interest.  They range from computing direct hydrodynamical
quantities, such as temperatures and densities, to derived quantities
like cooling times, two--body relaxation times, X--ray luminosities
and inertial tensors.  To study flattened objects such as galactic or
proto stellar disks versatile routines to find such objects and derive
projections, surface densities and other useful diagnostic quantities
were created or modified so that they understood the structure of the
hierarchy.

Visualization has proven equally involved. To allow interactive
exploration of the full data sets on available systems we developed
Jacques, a GUI-based visualization tool which allows simultaneous
interactive analysis of tens of thousands of grids of the AMR
hierarchy on modest memory machines.  Jacques is implemented in the
Interactive Data Language (IDL), and was used to produce the images in
Fig.~\ref{fig:density_zoom}.  The hundred solar radii protostar in
our computational volume is tiny compared to a needle in a haystack
and navigation techniques had to be devised to simplify the
identification of regions of interest and to visualize slices and
projections of them (Jacques has a ``zoom in by $10^{10}$ button''!).
Other features include velocity fields, isosurfaces, and a preliminary
volume renderer.

\section{Conclusions}

We have presented results on the highest-resolution 3-d computational
fluid dynamics simulation yet performed, achieving a spatial dynamic
range of $10^{12}$ in the region of interest.  The simulation breaks
new ground in the complexity of the method and the realism of the
result.  On the computational front, we solve a series of coupled
partial and ordinary differential equations with a range of
techniques, in parallel, on a hierarchy of 34 levels containing nearly
10,000 grids.  From an astrophysical point of view, we have performed
a long sought feat: the {\it ab initio} simulation of one of the first
stars in the universe.  Put together, we argue that this represents a
watershed in scientific computing.

\vspace{1cm}

We acknowledge support from NSF grant AST-9803137.  Support for GLB
was provided by NASA through Hubble Fellowship grant HF-01104.01-98A
from the Space Telescope Science Institute, operated by the
Association of Universities for Research in Astronomy, Inc., under
NASA contract NAS6-26555.  We gratefully acknowledge the support of
the San Diego Supercomputing Center.

\end{document}